\documentclass[aip,jcp,preprint,amsmath,amssymb,amsfonts,showkeys,showpacs,preprintnumbers]{revtex4-1}
\usepackage{bm,graphicx,xcolor,hyperref,lineno,multirow,microtype}

\DeclareMathOperator{\erf}{erf}
\DeclareMathOperator{\erfc}{erfc}

\newcommand{\ba}{\mathbf{a}}
\newcommand{\bb}{\mathbf{b}}
\newcommand{\bc}{\mathbf{c}}
\newcommand{\bd}{\mathbf{d}}
\newcommand{\be}{\mathbf{e}}
\newcommand{\br}{\mathbf{r}}

\newcommand{\bA}{\mathbf{A}}
\newcommand{\bB}{\mathbf{B}}
\newcommand{\bP}{\mathbf{P}}

\newcommand{\clm}{c_{lm}}
\newcommand{\Dj}{\widehat{D}_j}
\newcommand{\GAB}{G_{\rm AB}}
\newcommand{\mc}{\multicolumn}
\newcommand{\mJ}{\mathcal{J}}

\begin{document}

\title{Resolutions of the Coulomb operator VI: Computation of auxiliary integrals}

\author{Taweetham Limpanuparb}
\thanks{Corresponding author}
\email{taweetham.limpanuparb@anu.edu.au}
\author{Joshua W. Hollett}
\email{jhollett@rsc.anu.edu.au}
\author{Peter M. W. Gill}
\email{peter.gill@anu.edu.au}
\affiliation{Research School of Chemistry, 
Australian National University, Canberra, ACT 0200, Australia}
\date{\today}

\begin{abstract}
We discuss the efficient computation of the auxiliary integrals that arise when resolutions of two-electron operators (specifically, the Coulomb operator [T. Limpanuparb, A. T. B. Gilbert, and P. M. W. Gill, J. Chem. Theory Comput. {\bf 7}, 830 (2011)]  and the long-range Ewald operator [T. Limpanuparb and P. M. W. Gill, J. Chem. Theory Comput. {\bf 7}, 2353 (2011)]) are employed in quantum chemical calculations.  We derive a recurrence relation that facilitates the generation of auxiliary integrals for Gaussian basis functions of arbitrary angular momentum and propose a near-optimal algorithm for its use.
\end{abstract}

\keywords{Coulomb resolution; auxiliary integrals; recurrence relations; Cholesky decomposition; density fitting}
\pacs{02.70.-c, 31.15.-p}
\maketitle

\section{Introduction}
The Coulomb operator $1/r_{12}$ in electronic Hamiltonians is the ultimate source of all technical and computational difficulties in quantum chemistry.  If it were absent, the many-electron Schr\"odinger equation would immediately separate into one-electron equations and its solution would be straightforward, even for large systems.  Recognition of this has led to much research on the construction and investigation of methods that replace the Coulomb operator by approximations in which the two-electron interaction is, in one sense or another, ``factorized''.

One of the oldest approaches is to recognize that, when the two particles are far apart, their interaction can be modeled accurately by a multipole expansion.  Although this expansion originated in classical mechanics \cite{Legendre85, KelloggBook, Appel85, Rokhlin85, Greengard87}, it can also be exploited in quantum mechanical calculations and this led to the Continuous Fast Multipole Methods \cite{CFMM94, CFMM96, Scuseria96}.

In an alternative strategy, the electron density can be modeled in a relatively small auxiliary basis, leading to the so-called density fitting techniques \cite{Whitten73, Baerends73, Yanez78, Alsenoy88}.  It turns out that it is generally preferable to model the electric field \cite{Dunlap79, Hall80, Dunlap83, Vahtras93, ErfcRI05, Ahlrichs09} or potential \cite{Model92} of the density, rather than the density itself.  Cholesky factorization \cite{Beebe77, Koch03, Lindh07, Boman08, Pedersen09} is another member of the density-fitting family and, by decomposing the two-electron integral matrix, achieves a fit of the electric field of the electron density.

In a series of papers, \cite{Hackbusch06, Flad07, Hackbusch07, Khoromskij09, Flad10, Hackbusch11} Hackbusch and co-workers have designed schemes for constructing tensor factorizations of many-electron objects (including the Coulomb operator) and such techniques have recently yielded impressive results in Hartree-Fock \cite{Valeev11} and correlated \cite{Hackbusch11} calculations on a variety of small molecules.

Pursuing a different tack, other researchers have sought to explore the benefits of partitioning the Coulomb operator into a short-range-but-singular part and a long-range-but-smooth part.  This idea was introduced to theoretical chemistry by Ewald \cite{Ewald21} in 1921 but has enjoyed a renaissance during the past 17 years \cite{Panas95,Savin96, KWIK96, CASE96, CAP96, Split97, Eshort99, FTC02, Sirbu02, Izmaylov06}.  The Ewald partition
\begin{equation} \label{eq:Ewald}
	\frac{1}{r_{12}} = \frac{\erfc(\omega r_{12})}{r_{12}} + \frac{\erf(\omega r_{12})}{r_{12}}
\end{equation}
has become especially popular within DFT and a host of functionals have been devised \cite{CADFT96, Hirao01, Ernzerhof03, Yanai04, Tawada04, Toulouse04, Gerber05, Baer05, Toulouse06, Vydrov06, Heyd06, Krukau06, Sato07, Hirao07, Poisson08, Chai08_1, Chai08_2, Brothers08, Krukau08} to treat the short-range (\textit{i.e.}~the erfc) part of the operator.

In work related to that of the Hackbusch group, we have published a series of papers on resolutions of two-electron operators into sums of products of one-electron functions.  The first four papers\cite{RO08,Lag09,RRSE09,QuasiRO11} focus on the Coulomb operator and the fourth\cite{QuasiRO11} uses a Bessel identity \cite{Diego} to show that, if $r_1+r_2 < 2\pi$, \footnote{We measure $r_1$, $r_2$ and $r_{12}$ in atomic units and the system of interest must therefore be scaled to fit inside a sphere of radius $\pi$ a.u.} then
\begin{subequations} \label{eq:CoulRes}
\begin{gather}
	\frac{1}{r_{12}} = \sum_{nlm}^\infty \phi_{nlm}^*(\br_1)\phi_{nlm}(\br_2)						\\
	\phi_{nlm}(\br) = 2\sqrt{2-\delta_{n,0}} \ j_l(nr) Y_{lm}(\br)									\label{eq:phiCoul}
\end{gather}
\end{subequations}
where $\delta_{ij}$ is the Kronecker delta function, $j_l$ is a spherical Bessel function and $Y_{lm}$ is a complex spherical harmonic. \cite{NISTbook}

The fifth paper in the series\cite{REwald11} focuses on the long-range part of the Ewald partition \eqref{eq:Ewald} and shows that
\begin{subequations} \label{eq:EwaldRes}
\begin{gather}
	\frac{\erf(\omega r_{12})}{r_{12}} = \sum_{nlm}^\infty \phi_{nlm}^*(\br_1)\phi_{nlm}(\br_2)	\\
	\phi_{nlm}(\br) = 4\sqrt{h_n \omega} \ j_l(2\eta_n \omega r) \,Y_{lm}(\br)					\label{eq:phiEwald}
\end{gather}
\end{subequations}
where the $h_n$ and $\eta_n$ are the weights and (positive) roots of standard Gauss-Hermite quadrature \cite{NISTbook}.  We have found that this resolution converges well, and is particularly effective for the calculation of long-range exchange energies.  When used in conjunction with a DFT functional for the short-range part of the Ewald partition it offers a potent computational tool.

In the present paper, we will assume that
\begin{equation} \label{eq:phiGen}
	\phi_{nlm}(\br) = q_n \,\mJ_l(\lambda_n r) \,R_{lm}(\lambda_n\br)
\end{equation}
where $\mJ_l(t) = t^{-l} j_l(t)$ and $R_{lm}(\br) = r^l Y_{lm}(\br)$ is a solid harmonic.\cite{HobsonBook, WhittakerBook}  Equations \eqref{eq:phiCoul} and \eqref{eq:phiEwald} are special cases of \eqref{eq:phiGen}.  For brevity, we suppress the index on $\lambda_n$ henceforth.

Resolutions \eqref{eq:CoulRes} and \eqref{eq:EwaldRes} convert the two-electron integral
\begin{equation}
	( \ba \bb | \bc \bd ) = \iint \chi_\ba^*(\br_1) \chi_\bb(\br_1) T(r_{12}) \chi_\bc^*(\br_2) \chi_\bd(\br_2) \,\mathrm{d}\br_1 \,\mathrm{d}\br_2
\end{equation}
into the sum
\begin{equation} \label{eq:abcdRes}
	( \ba \bb | \bc \bd ) = \sum_{nlm}^\infty ( \ba \bb | nlm ) ( nlm | \bc \bd )
\end{equation}
of products of one-electron auxiliaries
\begin{equation} \label{eq:abphi}
	( \ba \bb | nlm ) = \int \chi_\ba^*(\br) \chi_\bb(\br) \phi_{nlm}(\br) \,\mathrm{d}\br
\end{equation}
In practice, Eq.~\eqref{eq:abcdRes} is not used to calculate two-electron integrals explicitly;  rather, it is substituted into quantum chemical expressions and then facilitates a variety of cost-saving factorizations.  For example, in three of the earlier papers\cite{RO08,QuasiRO11, REwald11} in this series, we derived expressions for the Coulomb and exchange energies in terms of the $( \ba \bb | nlm )$ rather than the $( \ba \bb | \bc \bd )$.  Equation \eqref{eq:abcdRes} has the same formal structure as found in Cholesky decomposition or density-fitting schemes, and hence the integral factorization resulting from the resolutions offers similar computational benefits.  Moreover the resolution approach avoids the cost of an (explicit or implicit) Cholesky decomposition or an (explicit or implicit) density-fitting matrix inversion.

However, the usefulness of resolutions such as \eqref{eq:CoulRes} and \eqref{eq:EwaldRes} is critically dependent on the accuracy of truncated version of the sum \eqref{eq:abcdRes} and the ease with which the auxiliaries \eqref{eq:abphi} can be generated.  In investigating the first of these issues, we discovered that surprisingly aggressive truncations of resolutions provide chemically meaningful Coulomb, exchange, MP2, full CI, long-range Coulomb and long-range exchange energies\cite{RO08, Lag09, RRSE09, QuasiRO11, REwald11}.  In the fourth and fifth papers \cite{QuasiRO11,REwald11} of this series, we addressed the second issue by presenting explicit formulae for the integrals \eqref{eq:abphi} where $\chi_\ba$ and $\chi_\bb$ are $s$- or $p$-type Gaussians.  However, in order that the resolutions be universally applicable, it is essential that an algorithm be available for constructing integrals over Gaussians of arbitrary angular momentum.

The explicit approach\cite{REwald11,KaewThesis} rapidly becomes complicated and inefficient as the angular momentum grows and it is generally agreed that the optimal schemes for forming Gaussian integrals are recursive. \cite{Boys50,Dupuis76, McMurchie78, Obara86, HGP88, Braket90, Review94, Ishida00, Ahlrichs06, Komornicki11}  Since the auxiliaries \eqref{eq:abphi} consist of a mixture of Gaussians, spherical Bessel functions and spherical harmonics, the standard Gaussian recurrence relations (RRs) are not immediately applicable.  However, we have recently used Ahlrichs' method to find an 18-term RR for a general class of Gaussian integrals in phase space\cite{RR11} and we were optimistic that an analogous attack would yield an RR for the auxiliaries.

In Section II of this paper, the sixth in the series, we derive an RR that allows auxiliaries to be constructed recursively from simple ingredients.  In Section III, we discuss an algorithm for applying the RR with near-optimal efficiency in terms of memory accesses.  The efficiency and numerical stability of the proposed algorithm are discussed in Section IV.

\section{Recurrence Relation for Auxiliaries}
If the basis functions in \eqref{eq:abphi} are the real Cartesian Gaussians
\begin{gather}
	[ \ba | = (x - A_x)^{a_x} (y - A_y)^{a_y} (z - A_z)^{a_z} e^{-\alpha |\br -\bA|^2}	\\
	[ \bb | = (x - B_x)^{b_x} (y - B_y)^{b_y} (z - B_z)^{b_z} e^{- \beta  |\br -\bB|^2}
\end{gather}
where $\ba=(a_x,a_y,a_z)$ and $\bb=(b_x,b_y,b_z)$ are vectors of angular momentum quantum numbers,\cite{Review94} then an RR may be derived using the approach of Ahlrichs\cite{Ahlrichs06}.

We define $\bm{1}_j=(\delta_{jx},\delta_{jy},\delta_{jz})$ and the scaled derivative
\begin{equation}
	\Dj = \frac{1}{2\alpha} \frac{\partial}{\partial A_j}
\end{equation}
and the Boys recurrence relation\cite{Boys50}
\begin{equation} \label{eq:Boys}
	[ \ba + \bm{1}_j | = \Dj [ \ba | + \frac{a_j}{2\alpha} [ \ba-\bm{1}_j |	
\end{equation}
shows that a higher Gaussian can be formed from two lower ones.  By induction, there exists an operator $\widehat{O}(\ba)$ that generates $[ \ba |$ from an $s$-type Gaussian,\cite{Ahlrichs06} \textit{i.e.}
\begin{equation}
	[ \ba | = \widehat{O}(\ba) [ \bm{0} |
\end{equation}
and which commutes with $\Dj$.  The operator factorizes
\begin{equation}
	\widehat{O}(\ba) = \widehat{O}(a_x) \widehat{O}(a_y) \widehat{O}(a_z)
\end{equation}
and it can be shown that
\begin{equation}	\label{eq:comm}
	\widehat{O}(a_j) Z = Z \widehat{O}(a_j) + a_j z \,\widehat{O}(a_j-1)
\end{equation}
for any function $Z$ that is linear in $A_j$, {\it i.e.}
\begin{equation}
	\Dj Z = z	\quad \text{and} \quad \Dj z = 0
\end{equation}
These properties will be important later.

The $\phi_{nlm}$ do not depend on $\bA$ and $\bB$, and therefore
\begin{equation}
	[ \ba \bb | nlm ] = \widehat{O}(\ba) \widehat{O}(\bb) [ \bm{00} | nlm ]
\end{equation}
Combining this with \eqref{eq:Boys} yields
\begin{equation}	\label{eq:modBoys}
	[ (\ba +\bm{1}_j) \bb | nlm ] = \widehat{O}(\ba) \widehat{O}(\bb) \Dj [ \bm{00} | nlm ]	
														+ \frac{a_j}{2\alpha} [ (\ba-\bm{1}_j) \bb | nlm ]
\end{equation}
and reveals, as has been noted elsewhere,\cite{Obara86,RR11} that the recursive properties of Gaussian integrals depend critically on the first derivative of their fundamental integral.\cite{ssss91}

The fundamental integral here is given\cite{QuasiRO11,REwald11} by
\begin{equation} \label{eq:00nlm}
	[ \bm{00} | nlm ] = \GAB \,\mJ_l(\lambda P) \,R_{lm}(\lambda\bP)
\end{equation}
where
\begin{gather}
	\zeta = \alpha+ \beta					\\
	\bP	= (\alpha \bA + \beta \bB) / \zeta	\\
	\GAB = q_n (\pi/\zeta)^{3/2}  \exp(-\lambda^2/4\zeta-\alpha \beta |\bA-\bB|^2/\zeta)
\end{gather}
but, because of the derivative identity
\begin{equation}  \label{eq:Jderiv}
	\Dj \left[ \mJ_l(\lambda P) \right] = (-\lambda^2/2\zeta) P_j\,\mJ_{l+1}(\lambda P)
\end{equation}
it turns out to be convenient to generalize \eqref{eq:00nlm} and consequently \eqref{eq:modBoys} to
\begin{equation} \label{eq:00nlmp}
	[ \bm{00} | nlm ]^{(p)} = (-\lambda^2/2\zeta)^p \GAB \,\mJ_{l+p}(\lambda P) \,R_{lm}(\lambda \bP)
\end{equation}
\begin{equation}	\label{eq:modBoysp}
	[ (\ba +\bm{1}_j) \bb | nlm ]^{(p)} = \widehat{O}(\ba) \widehat{O}(\bb) \Dj [ \bm{00} | nlm ]^{(p)}	
														+ \frac{a_j}{2\alpha} [ (\ba-\bm{1}_j) \bb | nlm ]^{(p)}
\end{equation}

Using the Gaussian derivative 
\begin{equation}
	\Dj \left[ \GAB \right] = (P_j - A_j) \,\GAB
\end{equation}
the $\mJ$ derivative \eqref{eq:Jderiv} and the solid harmonic derivatives
\begin{subequations}
\begin{align}
	\partial_x R_{lm}	& = \clm^- \, R_{l-1,m+1} - \clm^+ \, R_{l-1,m-1}	\\
	\partial_y R_{lm}	& = -i\left(\clm^- \, R_{l-1,m+1} + \clm^+ \, R_{l-1,m-1}\right)	\\
	\partial_z R_{lm}	& = \clm \, R_{l-1,m}
\end{align}
\end{subequations}
where
\begin{subequations}
\begin{gather}
	\clm^{\pm}	=	\sqrt{\frac{(l \pm m -1)(l \pm m)}{4}\frac{2l+1}{2l-1}}	\\
	\clm		=	\sqrt{(l^2-m^2)\frac{2l+1}{2l-1}}
\end{gather}
\end{subequations}
one can then show that
\begin{multline} \label{eq:D00nlmp}
	\Dj [ \bm{00} | nlm]^{(p)} = (P_j - A_j) [ \bm{00} | nlm ]^{(p)} + P_j [ \bm{00} | nlm ]^{(p+1)}													\\
								- \frac{\delta_{j,x}}{\lambda} (\clm^- [ \bm{00} | n,l-1,m+1 ]^{(p+1)} - \clm^+ [ \bm{00} | n,l-1,m-1 ]^{(p+1)})		\\
								+ \frac{i \delta_{j,y}}{\lambda} (\clm^- [ \bm{00} | n,l-1,m+1 ]^{(p+1)} + \clm^+ [ \bm{00} | n,l-1,m-1 ]^{(p+1)})		\\
								- \frac{\delta_{j,z}}{\lambda} \clm [ \bm{00} | n,l-1,m ]^{(p+1)}
\end{multline}
and the $[ \bm{00} | nlm]^{(p)}$ are thus closed under differentiation.

Substituting \eqref{eq:D00nlmp} into \eqref{eq:modBoysp} and applying the commutation relation \eqref{eq:comm} for $\widehat{O}(\ba)$ and $\widehat{O}(\bb)$ produces four new integrals of lower angular momentum because the $(P_j-A_j)$ and $P_j$ prefactors are linear in $A_j$ and $B_j$.  The commutation relation introduces no additional integrals for the remaining terms, as their prefactors do not depend on $A_j$ or $B_j$.  This finally yields the 8-term (for $j = x$ or $y$) or 7-term (for $j = z$) RR
\begin{multline} \label{eq:VRR}
	[ (\ba +\bm{1}_j) \bb | nlm ]^{(p)} =  (P_j - A_j) [ \ba \bb | nlm ]^{(p)} + P_j [ \ba \bb | nlm ]^{(p+1)}										\\
									+ \frac{a_j}{2\zeta} \left\{ [ (\ba -\bm{1}_j) \bb | nlm ]^{(p)} + [ (\ba -\bm{1}_j) \bb | nlm ]^{(p+1)} \right\}	\\
									+ \frac{b_j}{2\zeta} \left\{ [ \ba (\bb -\bm{1}_j) | nlm ]^{(p)} + [ \ba (\bb -\bm{1}_j) | nlm ]^{(p+1)} \right\}	\\
								- \frac{\delta_{j,x}}{\lambda} (\clm^- [ \ba \bb | n,l-1,m+1 ]^{(p+1)} - \clm^+ [ \ba \bb | n,l-1,m-1 ]^{(p+1)})		\\
								+ \frac{i \delta_{j,y}}{\lambda} (\clm^- [ \ba \bb | n,l-1,m+1 ]^{(p+1)} + \clm^+ [ \ba \bb | n,l-1,m-1 ]^{(p+1)})	\\
															- \frac{\delta_{j,z}}{\lambda} \clm [ \ba \bb | n,l-1, m  ]^{(p+1)}
\end{multline}
which allows us to generate $[ \ba \bb | nlm ]$ integrals from the fundamental integrals \eqref{eq:00nlmp}.  We also note that \eqref{eq:VRR} cannot be used if $\lambda=0$.  However, in this special case, the auxiliaries are trivial.

For clarity, we have derived the RR \eqref{eq:VRR} in terms of complex solid harmonics $R_{lm}$.  However, it is straightforward to construct the equivalent RR for real solid harmonics and, for optimal computational efficiency, this is preferable to the complex form.  Because the structures of the real and complex RRs are identical, the discussion in the following sections is applicable to both cases.

\section{Algorithm for auxiliary evaluation} \label{sec:alg}
Possessing an RR that relates complicated $[ \ba \bb | nlm ]$ integrals to simpler ones is a good start, but it does not constitute a computational pathway until one specifies precisely how it is to be used.  We now discuss a simple algorithm whose efficiency is close to optimal.

We define an $[ ab ]^{(p)}$ class as the set of all integrals that arise when two Cartesian Gaussian shells, with angular momenta $a$ and $b$, are combined with all the $\phi_{nlm}$ with $0 \le n \le N$, $0 \le l \le L$ and $-l \le m \le +l$.  If the superscript $p$ is missing, it implies $p=0$.

The $\phi_{nlm}$ do not appear explicitly in our class notation and this reflects the fact that all steps in the algorithm must be performed \emph{for every} $\phi_{nlm}$.  From this point, therefore, we will refer only to the bras of the integrals and the reader must remember that our statements refer implicitly to the full set of $(N+1)(L+1)^2$ kets.

Suppose that we wish to compute an $(ab)$ class, where the round brackets indicate a contracted class.  If we think retrosynthetically \cite{Corey88} and recall that the Head-Gordon and Pople ``horizontal RR'' (HRR)
\begin{equation} \label{eq:HRR}
	( \ba (\bb +\bm{1}_j) | = ( (\ba +\bm{1}_j) \bb | + (A_j -B_j) ( \ba \bb |
\end{equation}
can be applied to contracted integrals,\cite{HGP88} it becomes clear that it is best to form the $(ab)$ class from $(e0)$ classes, where $e = a, a+1, \ldots, a+b$.  This tactic is beneficial because, when constructing classes with $b=0$, the fifth and sixth terms in the RR \eqref{eq:VRR} vanish.

Next, we note that the RR forms the $[(e+1)0]^{(p)}$ class from the $[e0]^{(p)}$ and $[(e-1)0]^{(p)}$ classes (where $p$ is unchanged) along with the $[e0]^{(p+1)}$ and $[(e-1)0]^{(p+1)}$ classes (where $p$ is incremented).  This suggests an algorithm in which the $[e0]^{(p)}$ are formed in an outer loop over $p = a+b,\ldots,0$ and an inner loop over $e = 0,\ldots,a+b-p$.  The $[00]^{(p)}$ classes are constructed using \eqref{eq:00nlmp}.

\begin{table}
	\caption{\label{tab:ddinterm} Intermediate classes needed to form a $(dd)$ class}
	\begin{ruledtabular}
	\begin{tabular}{cccccccc}
																		\mc{8}{c}{Generation}														\\
				V0		&		V1			&		V2			&		V3			&		V4		&		C		&		H1		&		H2		\\
		\hline
		$[00]^{(4)}$		&	$[00]^{(3)}$		&	$[00]^{(2)}$		&	$[00]^{(1)}$		&	$[00]$		&		---		&		---		&		---		\\
				---		&	$[10]^{(3)}$		&	$[10]^{(2)}$		&	$[10]^{(1)}$		&	$[10]$		&		---		&		---		&		---		\\
				---		&			---		&	$[20]^{(2)}$		&	$[20]^{(1)}$		&	$[20]$		&	$(20)$		&		---		&		---		\\
				---		&			---		&		---			&	$[30]^{(1)}$		&	$[30]$		&	$(30)$		&	$(21)$		&		---		\\
				---		&			---		&		---			&		---			&	$[40]$		&	$(40)$		&	$(31)$		&	$(22)$		\\
	\end{tabular}
	\end{ruledtabular}
\end{table}

Table \ref{tab:ddinterm} shows how the algorithm constructs a $(dd)$ class.  The $[00]^{(p)}$ class in each V generation is found via \eqref{eq:00nlmp}.  Each of the other V classes is formed, using \eqref{eq:VRR}, from classes above it in its own generation and the preceding one.  The C classes are formed by contracting those in the final V generation and, finally, the H classes are formed using \eqref{eq:HRR}.

Having determined the gross structure of the algorithm, \emph{i.e.}~which intermediate classes are needed, we next ask how each of these classes can be efficiently constructed.  Because modern computer architectures are limited primarily by data access, rather than floating-point operations,\cite{Mops93} we measure the cost of our algorithm in terms of fetch count.

\subsection{Construction of $[00]^{(p)}$ classes}
The $[00]^{(p)}$ classes are defined by \eqref{eq:00nlmp} but it is important to note that the $(-\lambda^2/2\zeta)$ and $\GAB$ do not depend on $l$ or $m$, the $\mJ_l(\lambda P)$ do not depend on $m$, and the $R_{lm}(\lambda \bP)$ depend only trivially on $n$.  It is therefore best to pre-compute these quantities and (if $N$ and $L$ are moderately large) this pre-computation will represent a negligible fraction of the total cost of forming an $(ab)$ class.

The $\mJ_l$ satisfy the backward recurrence relation \cite{NISTbook, Olver72}
\begin{equation}
	\mJ_l(t) = (2l+3) \mJ_{l+1}(t) - t^2 \mJ_{l+2}(t)								
\end{equation}
and the two initial values $\mJ_{L+a+b}(t)$ and $\mJ_{L+a+b-1}(t)$ can be obtained by calling a standard Bessel function routine\cite{Rlanguage}.

The solid harmonics satisfy the forward recurrence relation \cite{NISTbook, NR3rded}
\begin{equation} \label{eq:RlmRR}
	R_{lm} = \sqrt{\frac{4l^2-1}{l^2-m^2}} \,z \,R_{l-1,m} - \sqrt{\frac{2l+1}{2l-3} \frac{(l-1)^2-m^2}{l^2-m^2}} \, r^2 R_{l-2,m}
\end{equation}
and the reflection formula
\begin{equation} \label{eq:RlmReflec}
	R_{l,-m} = (-1)^m R_{lm}^*
\end{equation}
and take the boundary values
\begin{equation} \label{eq:RlmInit}
	R_{ll} = (-1)^l \sqrt{\frac{(2l+1)!!}{4\pi(2l)!!}}  \ (x+iy)^l
\end{equation}
Thus, it is efficient to construct the $R_{ll}$ ($0 \le l \le L$) using \eqref{eq:RlmInit}, form the $R_{lm}$ ($0 \le m < l$) via \eqref{eq:RlmRR} and the $R_{l,-m}$ by \eqref{eq:RlmReflec}. 

If the factors in \eqref{eq:00nlmp} have been thus pre-computed, the assembly of all the necessary $[00]^{(p)}$ classes requires only
\begin{equation} \label{eq:costA}
	F_1 = 3(a+b+1)
\end{equation}
memory fetches per ket.

\subsection{Construction of $[e0]^{(p)}$ classes}
A quick examination reveals that the RR \eqref{eq:VRR} is not equally costly for all integrals.  Specifically, it is cheapest when $a_j = 0$ (for then the third and fourth terms vanish) and when $j=z$ (for then the solid harmonic derivative involves only a single term).  Since it is desirable to minimize the number of integrals on the right-hand side of \eqref{eq:VRR} that must be fetched, we propose the following boolean for choosing the increment direction $j$ when using \eqref{eq:VRR} to form a member of the $[e0]^{(p)}$ class:
\begin{align*}
	\text{If }			e_z = 1,	&	\quad \text{choose }	j=z		&	\qquad	\text{(3 fetches)}	\\
	\text{else if }	e_y = 1,	&	\quad \text{choose }	j=y		&	\qquad	\text{(4 fetches)}	\\
	\text{else if }	e_x = 1,	&	\quad \text{choose }	j=x		&	\qquad	\text{(4 fetches)}	\\
	\text{else if }	e_z \ge 2,	&	\quad \text{choose }	j=z		&	\qquad	\text{(5 fetches)}	\\
	\text{else if }	e_y \ge 2,	&	\quad \text{choose }	j=y		&	\qquad	\text{(6 fetches)}	\\
	\text{else,}					&	\quad \text{choose }	j=x		&	\qquad	\text{(6 fetches)}
\end{align*}
Table \ref{tab:reduce} lists the best $j$ and associated fetch cost for all Cartesian Gaussians $[\be|$ with $e \le 6$.

\begin{table}
	\caption{\label{tab:reduce} Increment direction ($j$) and number of fetches ($f$) for each integral when forming an $[e0]^{(p)}$ class by \eqref{eq:VRR}}
	\begin{ruledtabular}  \scriptsize {
		\begin{tabular}{ccccccccc} 
			\mc{3}{c}{$e=1$ or 3 or 4}				&		\mc{3}{c}{$e=2$ or 5}				&		\mc{3}{c}{$e=6$}			\\
					\cline{1-3}									\cline{4-6}									\cline{7-9}
			$\be$		&	$j$		&	$f$			&	$\be$		&	$j$		&	$f$			&	$\be$		&	$j$		&	$f$	\\
			\hline
			$(1,0,0)$	&	$x$		&	4			&	$(2,0,0)$	&	$x$		&	6			&	$(6,0,0)$	&	$x$		&	6	\\
			$(0,1,0)$	&	$y$		&	4			&	$(1,1,0)$	&	$y$		&	4			&	$(5,1,0)$	&	$y$		&	4	\\
			$(0,0,1)$	&	$z$		&	3			&	$(1,0,1)$	&	$z$		&	3			&	$(5,0,1)$	&	$z$		&	3	\\
						&			&				&	$(0,2,0)$	&	$y$		&	6			&	$(4,2,0)$	&	$y$		&	6	\\
			$(3,0,0)$	&	$x$		&	6			&	$(0,1,1)$	&	$z$		&	3			&	$(4,1,1)$	&	$z$		&	3	\\
			$(2,1,0)$	&	$y$		&	4			&	$(0,0,2)$	&	$z$		&	5			&	$(4,0,2)$	&	$z$		&	5	\\
			$(2,0,1)$	&	$z$		&	3			&				&			&				&	$(3,3,0)$	&	$y$		&	6	\\
			$(1,2,0)$	&	$x$		&	4			&	$(5,0,0)$	&	$x$		&	6			&	$(3,2,1)$	&	$z$		&	3	\\
			$(1,1,1)$	&	$z$		&	3			&	$(4,1,0)$	&	$y$		&	4			&	$(3,1,2)$	&	$y$		&	4	\\
			$(1,0,2)$	&	$x$		&	4			&	$(4,0,1)$	&	$z$		&	3			&	$(3,0,3)$	&	$z$		&	5	\\
			$(0,3,0)$	&	$y$		&	6			&	$(3,2,0)$	&	$y$		&	6			&	$(2,4,0)$	&	$y$		&	6	\\
			$(0,2,1)$	&	$z$		&	3			&	$(3,1,1)$	&	$z$		&	3			&	$(2,3,1)$	&	$z$		&	3	\\
			$(0,1,2)$	&	$y$		&	4			&	$(3,0,2)$	&	$z$		&	5			&	$(2,2,2)$	&	$z$		&	5	\\
			$(0,0,3)$	&	$z$		&	5			&	$(2,3,0)$	&	$y$		&	6			&	$(2,1,3)$	&	$y$		&	4	\\
						&			&				&	$(2,2,1)$	&	$z$		&	3			&	$(2,0,4)$	&	$z$		&	5	\\
			$(4,0,0)$	&	$x$		&	6			&	$(2,1,2)$	&	$y$		&	4			&	$(1,5,0)$	&	$x$		&	4	\\
			$(3,1,0)$	&	$y$		&	4			&	$(2,0,3)$	&	$z$		&	5			&	$(1,4,1)$	&	$z$		&	3	\\
			$(3,0,1)$	&	$z$		&	3			&	$(1,4,0)$	&	$x$		&	4			&	$(1,3,2)$	&	$x$		&	4	\\
			$(2,2,0)$	&	$y$		&	6			&	$(1,3,1)$	&	$z$		&	3			&	$(1,2,3)$	&	$x$		&	4	\\
			$(2,1,1)$	&	$z$		&	3			&	$(1,2,2)$	&	$x$		&	4			&	$(1,1,4)$	&	$y$		&	4	\\
			$(2,0,2)$	&	$z$		&	5			&	$(1,1,3)$	&	$y$		&	4			&	$(1,0,5)$	&	$x$		&	4	\\
			$(1,3,0)$	&	$x$		&	4			&	$(1,0,4)$	&	$x$		&	4			&	$(0,6,0)$	&	$y$		&	6	\\
			$(1,2,1)$	&	$z$		&	3			&	$(0,5,0)$	&	$y$		&	6			&	$(0,5,1)$	&	$z$		&	3	\\
			$(1,1,2)$	&	$y$		&	4			&	$(0,4,1)$	&	$z$		&	3			&	$(0,4,2)$	&	$z$		&	5	\\
			$(1,0,3)$	&	$x$		&	4			&	$(0,3,2)$	&	$z$		&	5			&	$(0,3,3)$	&	$z$		&	5	\\
			$(0,4,0)$	&	$y$		&	6			&	$(0,2,3)$	&	$z$		&	5			&	$(0,2,4)$	&	$z$		&	5	\\
			$(0,3,1)$	&	$z$		&	3			&	$(0,1,4)$	&	$y$		&	4			&	$(0,1,5)$	&	$y$		&	4	\\
			$(0,2,2)$	&	$z$		&	5			&	$(0,0,5)$	&	$z$		&	5			&	$(0,0,6)$	&	$z$		&	5	\\
			$(0,1,3)$	&	$y$		&	4			&				&			&				&				&			&		\\
			$(0,0,4)$	&	$z$		&	5			&				&			&				&				&			&		\\ 
		\end{tabular}  }
	\end{ruledtabular}
	Totals for $e = 1,\ldots,6$ are $11,27,42,65,92,124$, respectively.
\end{table}

Because there are $\binom{e+2}{2}$ bras in an $[e0]^{(p)}$ class, the total fetch cost of forming all of these classes is
\begin{equation} \label{eq:costB}
	F_2 = \mu \sum_{p=0}^{a+b} \sum_{e=0}^{a+b-p} \binom{e+2}{2} = \mu \binom{a+b+4}{4}
\end{equation}
where $\mu$, the average fetch cost of \eqref{eq:VRR}, is roughly 4.

\subsection{Construction of $(e0)$ classes}
The reduction of primitive $[e0]$ classes into contracted $(e0)$ classes is achieved by the contraction step
\begin{equation}
	(e0) = \sum_{k_a=1}^{K_a} \sum_{k_b=1}^{K_b} D_{a,k_a} D_{b,k_b} [e0]
\end{equation}
where $K_a$ and $K_b$ are the degrees of contraction of $(\ba|$ and $(\bb|$, respectively, and $D_{a,k_a}$ and $D_{b,k_b}$ are the associated contraction coefficients.

The total fetch cost of forming all of the $(e0)$ classes from the $[e0]$ classes is
\begin{equation} \label{eq:costC}
	F_3 = \sum_{e=a}^{a+b} \binom{e+2}{2} = \binom{a+b+3}{3} - \binom{a+2}{3}
\end{equation}
 
\subsection{Construction of the $(ab)$ class}
The target $(ab)$ class can be formed from the $(e0)$ classes using either \eqref{eq:HRR} or the more general TR$n$ relations \cite{TRn93} but, for simplicity, we will consider only the former here.

Assuming that the internuclear distance $A_j-B_j$ (which is constant for all the kets) is pre-loaded, the horizontal RR requires that two integrals be fetched.  Using the fact that there are $\binom{e+2}{2}\binom{f+2}{2}$ bras in an $(ef)$ class, the total fetch cost of forming $(ab)$ from the $(e0)$ classes is
\begin{equation} \label{eq:costD}
	F_4 = 2 \sum_{f=1}^b \sum_{e=a}^{a+b-f} \binom{e+2}{2} \binom{f+2}{2}
\end{equation}
This can be written as an untidy polynomial which is 2nd degree in $a$ and 6th degree in $b$.

\section{Computational cost and numerical stability}

\subsection{Computational cost}

\begin{table}
	\caption{\label{tab:ddcost} Fetch costs to form a $(dd)$ class}
	\begin{ruledtabular}
		\begin{tabular}{cccccccc}
									\mc{6}{c}{Primitive}							&		\mc{2}{c}{Contracted}	\\
			V0		&	V1		&	V2		&	V3		&	V4		&	C		&		H1		&		H2		\\
										\cline{1-6}											\cline{7-8}
			3		&	3		&	3		&	3		&	3		&	---		&		---		&		---		\\
			---		&	11		&	11		&	11		&	11		&	---		&		---		&		---		\\
			---		&	---		&	27		&	27		&	27		&	6		&		---		&		---		\\
			---		&	---		&	---		&	42		&	42		&	10		&		36		&		---		\\
			---		&	---		&	---		&	---		&	65		&	15		&		60		&		72		\\
		\end{tabular}
	\end{ruledtabular}
\end{table}

The fetch cost for forming an $(ab)$ class is
\begin{equation}
	\text{Cost} = (F_1 + F_2 + F_3) K_a K_b + F_4
\end{equation}
The first three cost parameters are multiplied by $K_a K_b$ because they apply to each of the primitive bras.

The results in Table \ref{tab:reduce}, and \eqref{eq:costA}, \eqref{eq:costC} and \eqref{eq:costD} can be used to find the cost of constructing any $(ab)$ class of interest.  To illustrate this, the costs of each of the intermediate classes in Table \ref{tab:ddinterm} are shown in Table \ref{tab:ddcost}, and it follows from these that the cost parameters for a $(dd)$ class are $F_1 = 15$, $F_2 = 274$, $F_3 = 31$ and $F_4 = 168$.

The cost parameters for any class up to $(f\!f)$ can be found in the same way, and these are given in Table \ref{tab:costs}.  A cursory glance at this Table reveals that the construction of the fundamental integrals ($F_1$) and the contraction step ($F_3$) are much cheaper than the construction of the $[e0]^{(p)}$ classes ($F_2$), except when the angular momentum of the bra is low.  The cost of the $(e0) \rightarrow (ab)$ transformation ($F_4$) is significant for bras of high angular momentum but the fact that this work is performed on contracted, not primitive, bras means that the construction of the $[e0]^{(p)}$ via \eqref{eq:VRR} will be the computational bottleneck for most classes arising in typical quantum chemistry calculations.

\begin{table}
	\caption{\label{tab:costs} Cost parameters for forming various $(ab)$ classes}
	\begin{ruledtabular}
		\begin{tabular}{cccccc}
					&			&				\mc{4}{c}{Fetch cost parameters}				\\
																	\cline{3-6}
			Class	&	Size	&		$F_1$	&		$F_2$	&		$F_3$	&		$F_4$	\\
			\hline
			$(ss)$	&	1		&		3		&		0		&		1		&		0		\\
			$(ps)$	&	3		&		6		&		11		&		3		&		0		\\
			$(ds)$	&	6		&		9		&		49		&		6		&		0		\\
			$(fs)$	&	10		&		12		&		129		&		10		&		0		\\ [2mm]
			$(pp)$	&	9		&		9		&		49		&		9		&		18		\\
			$(dp)$	&	18		&		12		&		129		&		16		&		36		\\
			$(fp)$	&	30		&		15		&		274		&		25		&		60		\\ [2mm]
			$(dd)$	&	36		&		15		&		274		&		31		&		168		\\
			$(fd)$	&	60		&		18		&		511		&		46		&		270		\\ [2mm]
			$(f\!f)$	&	100		&		21		&		872		&		74		&		776		\\
		\end{tabular}
	\end{ruledtabular}
\end{table}

The structure of our RR \eqref{eq:VRR} is similar to that of the VRR of Head-Gordon and Pople\cite{HGP88} and, as a result, their fetch and FLOP costs are comparable.  Examination of Table \ref{tab:costs} reveals that the formation of a single, uncontracted auxiliary integral requires approximately 10 fetches, irrespective of its angular momentum.  This is comparable to the cost of forming a single, uncontracted conventional two-electron integral by the HGP algorithm\cite{HGP88} and implies that the use of truncated versions of \eqref{eq:abcdRes} will be competitive with traditional approaches provided that the number of resolution functions $\phi_{nlm}$ is less than the number of basis function pairs $\chi_\ba \chi_\bb$ in the system.

For calculations with heavily contracted basis sets, there exist algorithms \cite{Pople78, Review94} that introduce contraction early and these offer significantly enhanced efficiency.  Although we will not explore this extension here, it is straightforward to use the PRISM approach\cite{Review94} to construct a contracted version of \eqref{eq:VRR} and this offers a cheaper pathway to $(ab)$ classes where $K_a K_b \gg 1$.

\subsection{Numerical stability}
In the foregoing sections, we have derived an 8-term RR \eqref{eq:VRR} to construct auxiliary integrals of high angular momentum.  We have advocated that it be used in conjunction with the HRR, and we have carefully considered the computational cost of the resulting algorithm.  However, we have not addressed the important practical question of numerical stability.  After all, if our new RR is unstable, the fact that it is cheap will be of little real interest.  Unfortunately, although the stability characteristics of two-term recurrence relations\cite{Olver1964, Gautschi1967, Olver1972, Mattheij1976, Zahar1977} -- including the HRR\cite{Ishida93, Knizia11} -- have been studied extensively, the \emph{theoretical} numerical stabilities of many-term recurrence relations, such as the VRR and our RR \eqref{eq:VRR}, are less well understood.  Therefore, having little option, we have resorted to extensive \emph{numerical} experiments.

We decided to focus on the construction of a $[gg]$ class, using the HRR to transfer angular momentum from $\bA$ to $\bB$.  We performed a high-dimensional scan, varying the Gaussian exponents ($\alpha,\beta = 10^{-2}, 10^{-1},\ldots, 10^5$), the intercenter distance ($R = |\bA-\bB| = 0, 1, 10, 100$), and the resolution parameters ($n = 0,1,10,100$, $l = 0, 5, 15, 25$, $m=0$, $q_n=1$), and using \eqref{eq:VRR} to generate the 45 $[ks|nlm]$ integrals, in both double and extended precision.  Remarkably, the largest absolute error in the double precision estimates of the integrals was $1.84\times10^{-16}$, indictating that the RR \eqref{eq:VRR} was completely stable for all of the 184 320 integrals investigated.  We conclude that the numerical stability of the new RR probably exceeds that of the widely used HRR\cite{Ishida93,Knizia11} and is, in any event, no cause for concern.

\section{Concluding Remarks}

In the present article, we have presented a detailed algorithm for the computational construction of the auxiliary integrals $( \ba \bb | nlm )$ and we have shown that, although such integrals involve Gaussians, Bessel functions and spherical harmonics, they can be generated recursively at approximately the same cost per integral as traditional two-electron Gaussian repulsion integrals.

This work opens the door to the routine and widespread application of two-electron resolutions within Gaussian-based quantum chemistry.  We are currently developing a high-performance implementation of our algorithm and we will report applications to problems of chemical interest in the near future.

\begin{acknowledgments}
P.M.W.G. thanks the ARC for funding (DP0984806 and DP1094170) and the NCI National Facility for supercomputer resources. 
\end{acknowledgments}

\bibliography{ROints}

\end{document}